\documentclass[preprint2]{aastex63}
\usepackage{epsfig}
\usepackage{amsmath}

\shortauthors{J. Li et al.}
\shorttitle{X-ray Brightest Quasar at z>5}

\begin{document}

\title{Chandra Detection of Three X-ray Bright Quasars at $z>5$}

\author[0000-0001-6239-3821]{Jiang-Tao Li}
\affiliation{Department of Astronomy, University of Michigan, 311 West Hall, 1085 S. University Ave, Ann Arbor, MI, 48109-1107, U.S.A.}

\author[0000-0002-7633-431X]{Feige Wang}
\altaffiliation{NHFP Hubble Fellow}
\affiliation{Steward Observatory, University of Arizona, 933 North Cherry Avenue, Tucson, AZ 85721, USA}

\author[0000-0001-5287-4242]{Jinyi Yang}
\altaffiliation{Strittmatter Fellow}
\affiliation{Steward Observatory, University of Arizona, 933 North Cherry Avenue, Tucson, AZ 85721, USA}

\author{Yuchen Zhang}
\affiliation{Department of Astronomy, University of Michigan, 311 West Hall, 1085 S. University Ave, Ann Arbor, MI, 48109-1107, U.S.A.}

\author{Yuming Fu}
\affiliation{Department of Astronomy, School of Physics, Peking University, Beijing 100871, China}

\author[0000-0002-1620-0897]{Fuyan Bian}
\affiliation{European Southern Observatory, Alonso de C\'ordova 3107, Casilla 19001, Vitacura, Santiago 19, Chile}

\author[0000-0001-6239-3821]{Joel N. Bregman}
\affiliation{Department of Astronomy, University of Michigan, 311 West Hall, 1085 S. University Ave, Ann Arbor, MI, 48109-1107, U.S.A.}

\author[0000-0003-3310-0131]{Xiaohui Fan}
\affiliation{Steward Observatory, University of Arizona, 933 North Cherry Avenue, Tucson, AZ 85721, USA}

\author[0000-0002-3119-9003]{Qiong Li}
\affiliation{Kavli Institute for Astronomy and Astrophysics, Peking University, Beijing 100871, China}
\affiliation{Tsinghua Center of Astrophysics \& Department of Astronomy, Tsinghua University, Beijing 100084, China}
\affiliation{Department of Astronomy, University of Michigan, 311 West Hall, 1085 S. University Ave, Ann Arbor, MI, 48109-1107, U.S.A.}

\author[0000-0002-7350-6913]{Xue-Bing Wu}
\affiliation{Kavli Institute for Astronomy and Astrophysics, Peking University, Beijing 100871, China}
\affiliation{Department of Astronomy, School of Physics, Peking University, Beijing 100871, China}

\author{Xiaodi Yu}
\affiliation{Department of Astronomy and Institute of Theoretical Physics and Astrophysics, Xiamen University, Xiamen, Fujian 361005, China}

\correspondingauthor{Jiang-Tao Li}
\email{pandataotao@gmail.com}

\begin{abstract}
We report \emph{Chandra} detection of three UV bright radio quiet quasars at $z\gtrsim5$. We have collected a sufficient number of photons to extract an X-ray spectrum of each quasar to measure their basic X-ray properties, such as the X-ray flux, power law photon index ($\Gamma$), and optical-to-X-ray spectral slope ($\alpha_{\rm OX}$). J074749+115352 at $z=5.26$ is the X-ray brightest radio-quiet quasar at $z>5$. It may have a short timescale variation (on a timescale of $\sim3800\rm~s$ in the observer's frame, or $\sim600\rm~s$ in the rest frame) which is however largely embedded in the statistical noise. We extract phase folded spectra of this quasar. There are two distinguishable states: a ``high soft'' state with an average X-ray flux $\sim2.7$ times of the ``low hard'' state, and a significantly steeper X-ray spectral slope ($\Gamma=2.40_{-0.32}^{+0.33}$ vs $1.78_{-0.24}^{+0.25}$). We also compare the three quasars detected in this paper to other quasar samples. We find that J074749+115352, with a SMBH mass of $M_{\rm SMBH}\approx1.8\times10^9\rm~M_\odot$ and an Eddington ratio of $\lambda_{\rm Edd}\approx2.3$, is extraordinarily X-ray bright. It has an average $\alpha_{\rm OX}=-1.46\pm0.02$ and a 2-10~keV bolometric correction factor of $L_{\rm bol}/L_{\rm2-10keV}=42.4\pm5.8$, both significantly depart from some well defined scaling relations. We compare $\Gamma$ of the three quasars to other samples at different redshifts, and do not find any significant redshift evolution based on the limited sample of $z>5$ quasars with reliable measurements of the X-ray spectral properties.
\end{abstract}

\keywords{high-redshift --- quasars: observations --- early universe  }

\section{Introduction} \label{sec:Intro}

Many supermassive black halos (SMBHs) located at the center of galaxies gained a significant fraction of their mass at high redshift. In particular, there are two quasars at $z>7.5$ with billion solar mass black halos (BHs) detected \citep{Banados18,Yang20}. There are also some extremely massive SMBHs detected at $z>6$ (\citealt{Wu15}), with a mass comparable to the most massive ones in the local Universe ($M_{\rm SMBH}\gtrsim10^{10}\rm M_\odot$, e.g., \citealt{McConnell11,vandenBosch12}). The existence of such massive BHs at such a small age of the Universe ($<1\rm~Gyr$) is challenging to the theory of the growth of SMBHs and their coevolution with the host galaxies (e.g., \citealt{Smidt18}). 

X-ray observations provide important information on the inner accretions disk and hot corona close to the SMBH. There are many studies of the redshift evolution of the accretion physics of SMBHs based on X-ray measurements of some key parameters (e.g., the X-ray photon index $\Gamma$ or the spectral slope between the rest frame UV and X-ray bands $\alpha_{\rm OX}$) or some well defined scaling relations (such as the relation between the rest frame UV luminosity $L_{\rm 2500\AA}$ and $\alpha_{\rm OX}$). The major conclusion based on some very limited samples (small sample size and/or poor X-ray data) is the accretion physics does not show significant redshift dependence up to $z\gtrsim7$ where the most distant quasars have been detected (e.g., \citealt{Nanni17,Pons19,Salvestrini19,Vito19,Wang20b}).

Quasars at $z>5$ typically have a low X-ray flux of $<10^{-14}\rm~ergs~s^{-1}~cm^{-2}$. Their X-ray properties are thus often poorly constrained or estimated based on a few assumptions (e.g., $L_{\rm X}$ estimated from the broad band counts rate assuming a fixed $\Gamma$; e.g., \citealt{Li20,Wang20b}). In particular, X-ray timing analyses of AGNs provide critical information on the geometry of the accretion disk and corona, as well as the emitting mechanisms close to the SMBH (e.g., \citealt{GonzalezMartin12,Jin20a,Jin20b}). However, due to the poor counting statistic, X-ray timing analyses of high-$z$ quasars are often difficult, and most of the existing timing analyses are on long timescales between different observations (e.g., \citealt{Timlin20}). We therefore need higher quality X-ray data to directly measure the X-ray properties of some well defined examples.

In this paper, we present new \emph{Chandra} observations of three UV bright quasars at $z\gtrsim5$. This redshift marks the end of the earliest fast growth stage of some most massive SMBHs at higher redshifts, as the average SMBH growth rate seems to slow down at lower redshifts (e.g., \citealt{Willott10,Trakhtenbrot11}). A comparison of X-ray observations of quasars at this redshift to the small number of existing high-quality X-ray observations of quasars at lower and higher redshifts will thus help us to understand how the accretion of SMBHs evolve over cosmic time. 

The present paper is organized as follows: In \S\ref{Sec:ObsDataReduc}, we present the basic reduction and spectral analysis of our \emph{Chandra} data. For the X-ray brightest quasar in our sample, J074749+115352, we also conduct the first timing analysis on a few ks timescales (in the observer's frame) for an object at such high redshift. In \S\ref{Sec:Discussion}, we estimate the SMBH mass of J074749+115352 based on the \ion{Mg}{2} line in a near-IR spectrum, followed by discussions on its spectral-timing properties, as well as a comparison of the three quasars studied in this paper to other quasar samples. Our conclusions are summarized in \S\ref{Sec:Summary}. Throughout the paper, we adopt a flat cosmology model with $H_{\rm 0}=70\rm~km~s^{-1}~Mpc^{-1}$, $\Omega_{\rm M}=0.3$, $\Omega_{\rm \Lambda}=0.7$, and $q_{\rm 0}=-0.55$. All the errors computed in this paper are quoted at 1~$\sigma$ confidence level.

\section{Observations and Data Analysis} \label{Sec:ObsDataReduc}

The three quasars studied in this paper are selected from a large $z\sim5$ UV-bright quasar sample developed in a few papers \citep{Wang16,Yang16,Yang17}. We selected the most luminous ($M_{\rm 1450}\lesssim-28$) radio quiet quasars for follow-up \emph{Chandra} observations, which was approved in Cycle~21 (PI: Li) and taken from Oct. 2019 to Jan. 2020. Basic quasar properties and the corresponding \emph{Chandra} observation information are summarized in Table~\ref{table:sample}.


\begin{figure*}
\begin{center}
\epsfig{figure=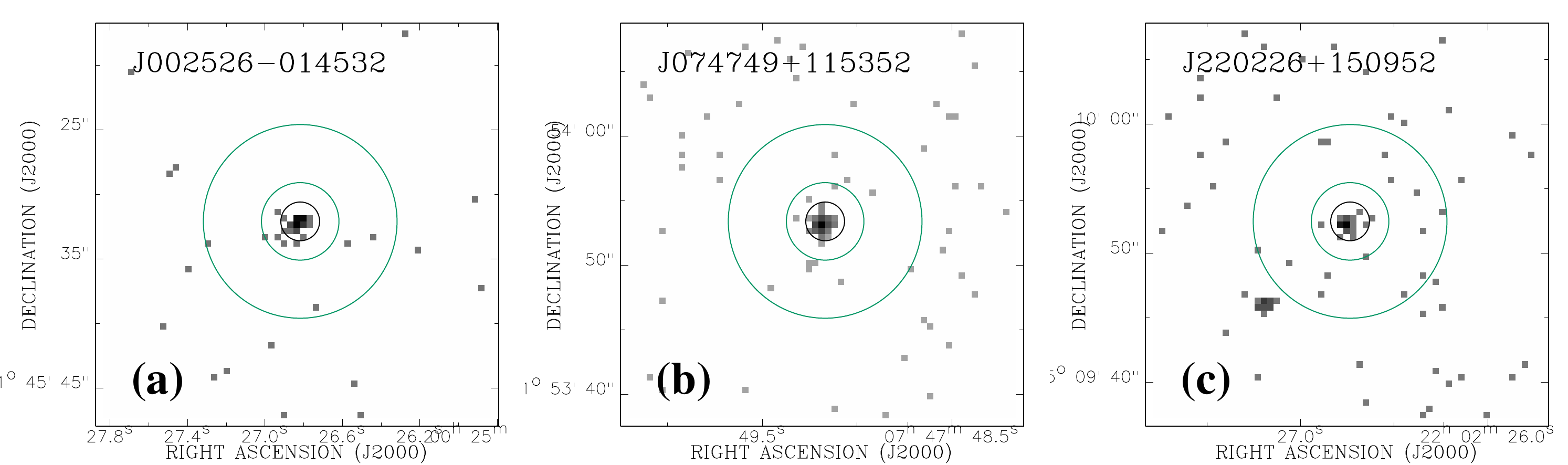,width=1.0\textwidth,angle=0, clip=}
\caption{0.5-7~keV \emph{Chandra} counts images of the three quasars studied in this paper, with their names denoted on top left of each panel. The small circle and larger annulus in each panel are the regions used to extract the source and background spectra (Fig.~\ref{fig:Chandraspec}), as well as the light curve and period folded spectra of J074749+115352 (Fig.~\ref{fig:J074749lc}), respectively.}\label{fig:Chandraimg}
\end{center}
\end{figure*}

\begin{figure*}
\begin{center}
\epsfig{figure=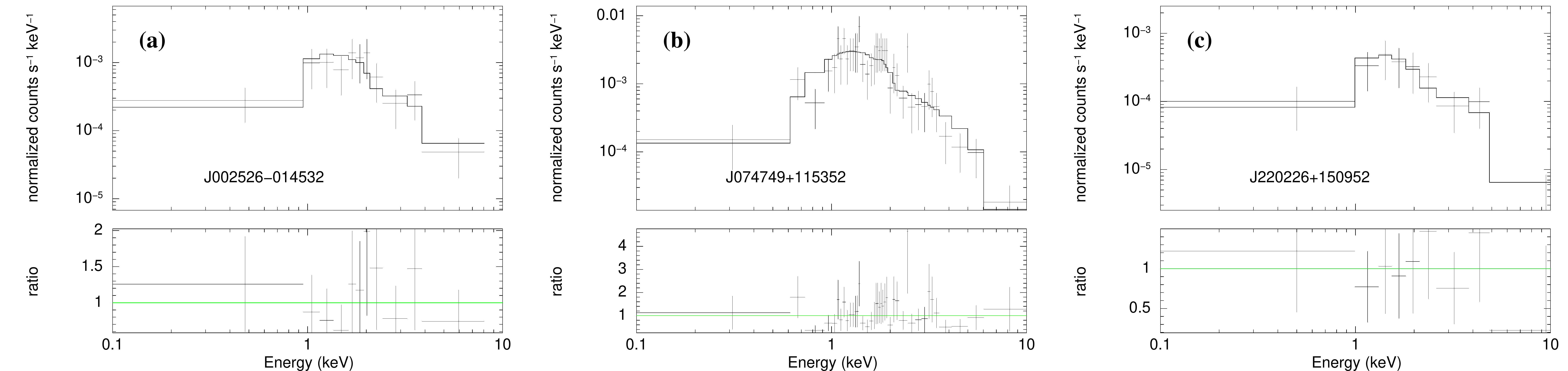,width=1.0\textwidth,angle=0, clip=}
\caption{The \emph{Chandra} spectrum of each quasar fitted with a power law model subjected to the Milky Way foreground extinction ($N_{\rm H}$ in Table~\ref{table:sample}). The minimum counts number in each bin is three.}\label{fig:Chandraspec}
\end{center}
\end{figure*}

The quasars in all of the \emph{Chandra} observations (in imaging mode) are located on ACIS-S3. We reduce the data in a uniform manner with CIAO~v4.11 and CALDB~v4.8.5. We reprocess the evt1 file in a standard way using the CIAO tool \emph{chandra\_repro}. We then define a $r=5^{\prime\prime}$ circular region centered at the initial optical position of the quasar, and compute the centroid position of the full band (0.5-7~keV) unbinned X-ray image. This step is conducted iteratively with smaller radius of the circular region ($r=3^{\prime\prime}$ and $2^{\prime\prime}$), in order to accurately determine the X-ray position of the object. We then extract the source and background spectra with the CIAO tool \emph{specextract} from a $r=1.5^{\prime\prime}$ circular region and a $r=3^{\prime\prime}-7.5^{\prime\prime}$ annulus centered at the X-ray source position computed above (Fig.~\ref{fig:Chandraimg}). In order to characterize the overall shape of the broad band X-ray spectra with the limited number of counts, we regroup the spectra with at least three counts in each bin (Fig.~\ref{fig:Chandraspec}). The spectra are analyzed with XSpec~v12.9.1. We adopt the Cash statistic and assume a redshifted power law model \emph{zpowerlw} subjected to Galactic foreground absorption described with the model \emph{tbabs} at a column density of $N_{\rm H}$ as listed in Table~\ref{table:sample}. At high redshifts (e.g., $z\sim5$ for quasars studied here), the observed X-ray photons at $>0.5\rm~keV$ correspond to a rest frame energy of $>3\rm~keV$, where the intrinsic absorption is typically negligible. We do not find any significant evidence for additional absorption, so do not include such a component in our spectral analysis. The intrinsic extinction corrected X-ray flux and the corresponding errors are calculated with the convolution model \emph{cflux}. We list the X-ray properties of the quasars obtained from our analysis in Table~\ref{table:XrayProperties}. When computing the rest frame UV (at $2500\rm~\AA$) to X-ray (at $2\rm~keV$) spectral index $\alpha_{\rm OX}$ from the absolute magnitude at $1450\rm~\AA$ ($M_{\rm 1450}$) and the measured full band X-ray luminosity, we adopt a UV spectral index of $\alpha=-0.5$ (a common assumption in similar studies, e.g., \citealt{Nanni17,Wang20b}) and the measured X-ray photon index $\Gamma$ (Table~\ref{table:XrayProperties}). The measured soft X-ray flux $F_{\rm 0.5-2 keV}$ is then converted to the rest frame $2-10\rm~keV$ luminosity $L_{\rm 2-10 keV}$.
 
\begin{table*}[]{}
\begin{center}
\small\caption{Properties of the selected most luminous $z=5-6$ quasars and their \emph{Chandra} observations}
\tabcolsep=6.5pt
\begin{tabular}{lccccccccccccccc}
\hline \hline
Name        & $z$ & $M_{\rm 1450}$ & $L_{\rm 2500 \AA}$ & $N_{\rm H}$                  &  ObsID & Obs Date & $t_{\rm exp}$ \\
            	&              & mag                   & $10^{32}\rm~erg~s^{-1}~Hz^{-1}$ & $10^{20}\rm~cm^{-2}$ &  & & ks \\
\hline
J002526.84-014532.51 & 5.07  & -28.70 & 1.72 & 2.92 & 22586 & 2019-10-30 & 14.58 \\
J074749.18+115352.46 & 5.26 & -28.04 & 0.94 & 3.83 & 22587 & 2020-01-05 & 29.58 \\
J220226.77+150952.38 & 5.07 & -28.02 & 0.92 & 5.19 & 22588 & 2019-09-19 & 27.90 \\
\hline \hline
\end{tabular}\label{table:sample}
\end{center}
Properties of the quasars are obtained from \citet{Wang16}. The redshifts are measured with the available UV and optical emission lines from high quality ground-based spectra. $M_{\rm 1450}$ is the absolute magnitude at $1450\rm~\AA$. $L_{\rm 2500 \AA}$ is the monochromatic luminosity at $2500\rm~\AA$, and is computed from $M_{\rm 1450}$ assuming a power law spectral index of $\alpha=-0.5$ in UV band. $N_{\rm H}$ is the Galactic foreground absorption column density toward the quasar obtained from the HEASARC webtools using the HI4PI full-sky \ion{H}{1} map \citep{HI4PI15}. The last three columns are the observational ID, start date, and effective exposure time after data calibration ($t_{\rm exp}$) of the \emph{Chandra} observations.
\end{table*}

All of the three quasars have been firmly detected at $>3~\sigma$ confidence level in both the soft ($0.5-2\rm~keV$) and hard ($2-7\rm~keV$) X-ray bands (Table~\ref{table:XrayProperties}; Fig.~\ref{fig:Chandraimg}). The location of the X-ray source well matches the optical counterpart, with an offset $\leq0.5^{\prime\prime}$ for each quasar. This offset is smaller than the angular resolution of \emph{Chandra}. It is thus unlikely to mis-identify the quasar with a foreground object. We have collected $>20$ net counts (after subtracting the local background) from each quasar which allows us to extract a spectrum for each of them (Fig.~\ref{fig:Chandraspec}). The spectra of all of the three quasars can be fitted with a power law whose photon index can be well constrained. The rest frame 2-10~keV luminosities of all of the quasars are $L_{\rm 2-10 keV}>10^{45}\rm~erg~s^{-1}$ ($>10^{46}\rm~erg~s^{-1}$ for J074749.18+115352.46; hereafter J074749+115352), much higher than any stellar X-ray sources in the host galaxy. We thus confirm a robust X-ray detection of the quasars with a negligible chance of mis-identification.

\begin{table*}[]{}
\begin{center}
\small\caption{Measured X-ray properties of the quasars}
\tabcolsep=3.5pt
\begin{tabular}{rccccccccccccccc}
\hline \hline
Name        &  $cts_{\rm S}$ & $cts_{\rm H}$ & $cts_{\rm F}$ &  $S/N_{\rm S}$ & $S/N_{\rm H}$ & $S/N_{\rm F}$ & $F_{\rm 0.5-2 keV}$ & $L_{\rm 2-10 keV}$ & $\Gamma$ & $\alpha_{\rm OX}$ \\ 
            	&     & & & & & & $10^{-14}\rm erg/s/cm^2$ & $10^{45}\rm erg~s^{-1}$ & & \\ 
\hline
J002526 & 19.8 & 13.0 & 32.8 & 4.4 & 3.6 & 5.7 & $1.41_{-0.38}^{+0.37}$ & $4.16_{-1.13}^{+1.08}$ & $1.80_{-0.33}^{+0.34}$ & $-1.70_{-0.05}^{+0.04}$ \\ 
\hline
J074749 & 92.9 & 43.8 & 136.7 & 9.6 & 6.6 & 11.7 & $3.49_{-0.45}^{+0.44}$ & $12.2\pm1.6$ & $2.07\pm0.17$ & $-1.46\pm0.02$ \\ 
High & 30.0 & 9.0 & 38.9 & 5.5 & 3.0 & 6.2 & $5.86_{-1.41}^{+1.36}$ & $22.9_{-5.5}^{+5.3}$ & $2.40_{-0.32}^{+0.33}$ & $-1.42\pm0.04$ \\ 
Low & 37.0 & 22.9 & 59.9 & 6.1 & 4.8 & 7.7 & $2.20_{-0.43}^{+0.42}$ & $6.95_{-1.38}^{+1.34}$ & $1.78_{-0.24}^{+0.25}$ & $-1.51\pm0.03$ \\ 
\hline
J220226 & 12.8 & 10.6 & 23.4 & 3.5 & 3.1 & 4.7 & $0.50_{-0.17}^{+0.16}$ & $1.41_{-0.47}^{+0.44}$ & $1.64_{-0.41}^{+0.42}$ & $-1.75_{-0.06}^{+0.05}$ \\ 
\hline \hline
\end{tabular}\label{table:XrayProperties}
\end{center}
``High'' and ``Low'' indicate the X-ray properties measured in the high soft and low hard states of J074749+115352, respectively. We list them together with the X-ray properties measured from the entire data of J074749+115352. The subscripts ``S'', ``H'', and ``F'' denote soft ($0.5-2\rm~keV$), hard ($2-7\rm~keV$), and full ($0.5-7\rm~keV$) band, respectively, all in the observer's frame. $cts$ is the net counts number after subtracting the local background. $S/N$ is the detection signal-to-noise ratio. $F_{\rm 0.5-2 keV}$ is the observed $0.5-2\rm~keV$ flux. $L_{\rm 2-10 keV}$ is the rest frame $2-10\rm~keV$ luminosity. $\Gamma$ is the X-ray photon index obtained from spectral fitting. $\alpha_{\rm OX}$ is the optical-to-X-ray spectral slope obtained from the rest frame monochromatic UV luminosity $L_{\rm 2500 \AA}$ and the measured monochromatic X-ray luminosity at $2\rm~keV$. For the ``High'' and ``Low'' states of J074749+115352, we use the same $L_{\rm 2500 \AA}$ to compute $\alpha_{\rm OX}$, which does not account for the variation of the UV flux, thus just listed for comparison.
\end{table*}

The observed $0.5-2\rm~keV$ flux of J074749+115352 is $F_{\rm 0.5-2 keV}=3.49_{-0.45}^{+0.44}\times10^{-14}\rm~erg~s^{-1}~cm^{-2}$. To our knowledge, this is \emph{the X-ray brightest radio quiet quasar at} $z>5$, only after two radio loud blazars [Q0906+6930 at $z=5.48$, \citet{Romani06}; SDSS J013127.34-032100.1 at $z=5.18$, \citet{An20}] and a radio loud non-blazar quasar CFHQS~J142952+544717 at $z=6.18$ recently detected in X-ray by \citet{Medvedev20} with \emph{eROSITA} (the X-ray brightest quasar at $z>5$, with an observed $0.3-2\rm~keV$ flux of $8.2_{-2.7}^{+3.7}\times10^{-14}\rm~erg~s^{-1}~cm^{-2}$). \citet{Medvedev20} have detected only nine counts (with an expected background contribution of $\approx0.8\rm~counts$) from a $r=30^{\prime\prime}$ circular region centered at the optical position of CFHQS~J142952+544717. For comparison, we have detected 136.7 net counts from a $r=1.5^{\prime\prime}$ region around J074749+115352, allowing us to perform \emph{the first timing analysis on a timescale of a few hours for an object at the cosmic dawn.}


\begin{figure*}
\begin{center}
\epsfig{figure=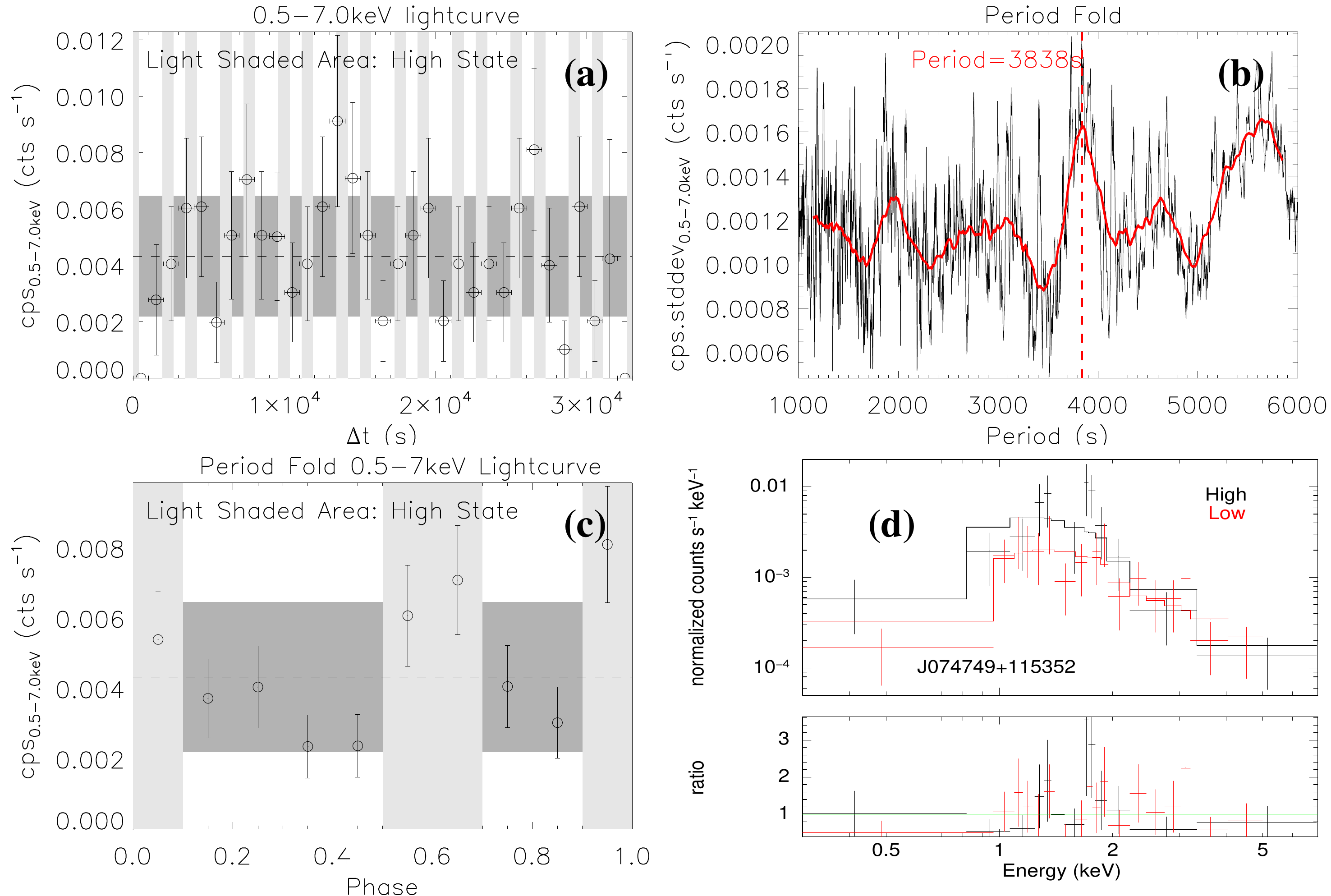,width=1.0\textwidth,angle=0, clip=}
\caption{Timing analysis of J074749+115352. All the parameters are measured in the observer's frame. (a) 0.5-7~keV light curve binned to a resolution of 1~ks. The dashed line and the dark shaded area are the mean value and standard deviation of the counts rate, respectively. The light shaded area marks the high soft state as analyzed in the other panels. (b) Period folded data to search for the most significant period of the X-ray counts rate variations. The $x$-axis is the assumed trial periods, while the $y$-axis is the standard deviation of the 0.5-7~keV counts rate of the period folded data. The black curve has been binned to a minimum step of 3~s, while the red curve is a smoothed version of the black curve. The red dashed line is the manually identified peak of the curve at a period of 3838~s. (c) Period folded 0.5-7~keV light curve at an assumed period of 3838~s. The light shaded areas identify the phase of the ``high state'', with a counts rate higher than the mean value (the dashed line and dark shaded area are the same as in panel a). (d) Spectra extracted from the ``high'' (black) and ``low'' (red) states identified in (c), respectively. The spectra are analyzed in the same way as those presented in Fig.~\ref{fig:Chandraspec}.}\label{fig:J074749lc}
\end{center}
\end{figure*}

For an accurate timing analysis of J074749+115352, we first apply a barycenter correction to the event, aspect solution (asol1), and exposure statistics (stat1) files, using the Level~1 orbit ephemeris file (eph1) created about one month after the observation (on 2020-02-04; see Table~\ref{table:sample} for the observation date). We then extract a $0.5-7\rm~keV$ light curve of J074749+115352 at a time resolution of $1\rm~ks$ (Fig.~\ref{fig:J074749lc}a). The standard deviation of the $0.5-7\rm~keV$ counts rate ($\approx2.1\times10^{-3}\rm~counts~s^{-1}$) is $\sim49\%$ of the mean counts rate ($\approx4.3\times10^{-3}\rm~counts~s^{-1}$). However, the variance of the light curve shown in Fig.~\ref{fig:J074749lc}a is even less than the mean square error, resulting in a negative excess variance \citep{Vaughan03}. This indicates the intrinsic variation is largely embedded in the large statistical noise. More sensitive X-ray observations are needed to confirm if a short time scale variation really exists or not. 

We make some further timing analysis of J074749+115352 to examine the possible time variation of the X-ray spectral properties. Limited by the poor counting statistics, we cannot obtain a high resolution light curve for a reliable Fourier power spectral analysis. In order to test the presence of any possible periodic signal and to search for the most significant variation period, we fold the event file with various trial periods using the CIAO tool \emph{pfold}. At the most significant period, there will be a local peak in the standard deviation of the counts rate. We search for such peaks in Fig.~\ref{fig:J074749lc}b, and find the most significant and narrow peak at a period of $P_{\rm obs}\approx 3838\rm~s$ in the observer's frame, which corresponds to a period of $P_{\rm rest}\approx 613\rm~s$ in the rest frame. We then create a period or phase folded light curve based on this period using the CIAO tool \emph{dmextract} (Fig.~\ref{fig:J074749lc}c). We identify $0.5\leq phase<0.7$, $\geq 0.9$, and $<0.1$ as the ``High'' state typically with a $0.5-7\rm~keV$ counts rate higher than the mean value, while the remaining phases as the ``Low'' state. The double peak probably indicates there is an unresolved shorter based period of $P_{\rm obs}\sim1900\rm~s$ which is however much weaker in Fig.~\ref{fig:J074749lc}b. We then create Good Time Intervals (GTIs) for each state with the CIAO tool \emph{dmgti} and align it with \emph{gti\_align}. We finally extract the spectra for each phase using \emph{specextract} and analyze them separately in the same way as adopted for the combined spectra above. The best-fit spectra are plotted together in Fig.~\ref{fig:J074749lc}d and the corresponding X-ray properties of the two phases are summarized in Table~\ref{table:XrayProperties}. We caution that we have adopted the same $L_{\rm 2500 \AA}$ to compute $\alpha_{\rm OX}$ for the ``High'' and ``Low'' states of J074749+115352.


\section{Discussions} \label{Sec:Discussion}

\subsection{An Estimate of the Mass of the SMBH in J074749+115352} \label{subsec:MSMBHJ074749}

\begin{figure}
\begin{center}
\includegraphics[width=0.48\textwidth]{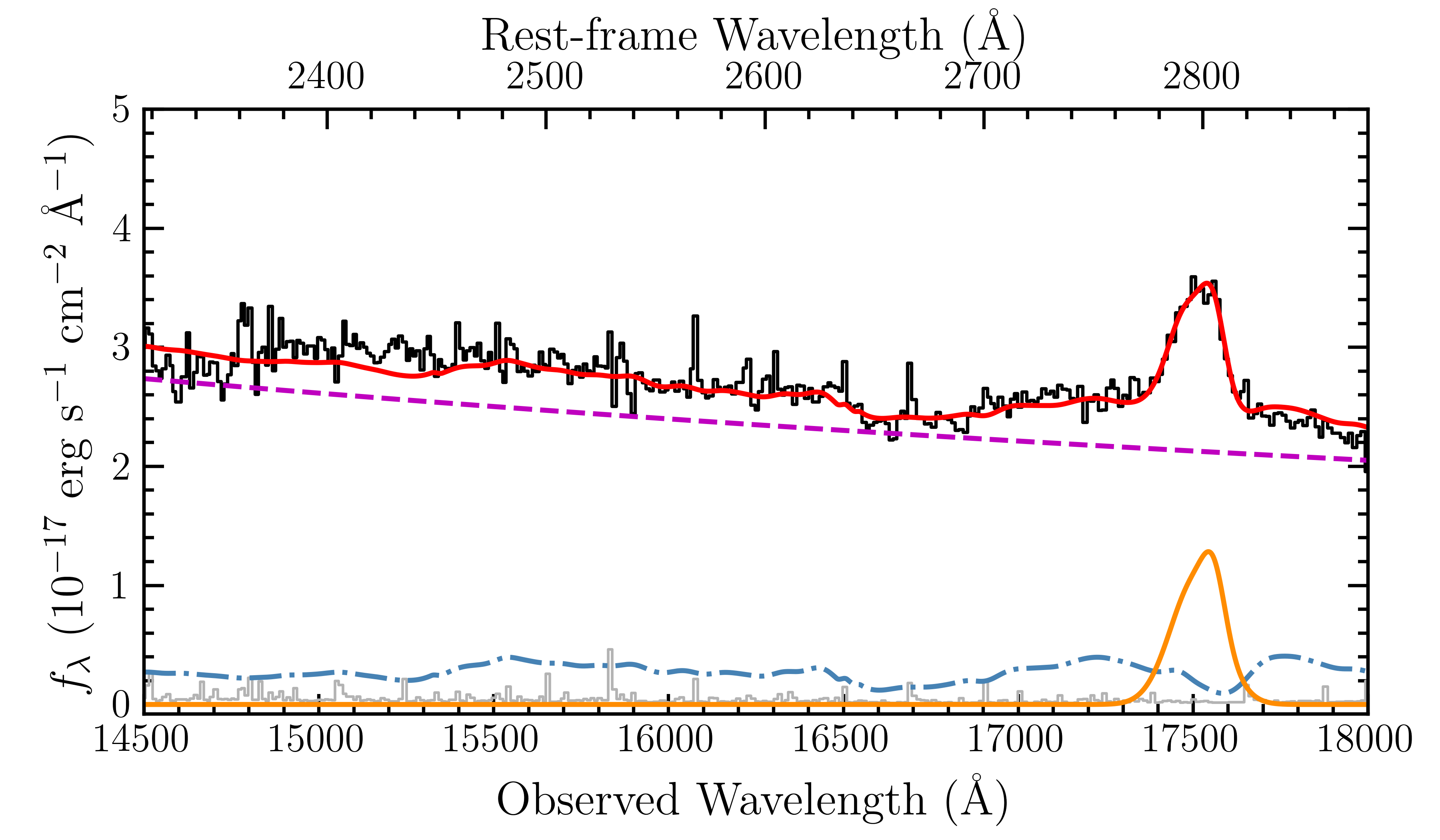}
\caption{Hale 5-m/TripleSpec H-band spectrum covering the \ion{Mg}{2}~$\lambda2800\rm~\AA$ emission line. The black and grey curves are the observed spectrum and the corresponding uncertainty, respectively. The purple dashed line, blue dash-dotted curve, and orange solid curve represent the best-fits of the power law continuum, the \ion{Fe}{2} template, and the \ion{Mg}{2} emission line, respectively. The red solid curve is a sum of the above best-fit model components.}\label{fig:J0747P200spec}
\end{center}
\end{figure}

We first estimate the mass of the SMBH ($M_{\rm SMBH}$) in J074749+115352 based on the \ion{Mg}{2}~$\lambda2800\rm~\AA$ emission line. The near-IR spectrum used in the measurement was taken in March, 2014 with the TripleSpec spectrograph on the Hale 5-m telescope at the Palomar observatory, with a $1^{\prime\prime}$ slit and a total exposure time of 6000~s (Fig.~\ref{fig:J0747P200spec}). We fit the spectrum with a pseudo-continuum, including a power-law continuum, the \ion{Fe}{2} template \citep{Tsuzuki06}, and the Balmer continuum \citep{DeRosa14}. The continuum-subtracted spectrum around the \ion{Mg}{2} line is fitted with a two-component Gaussian model. The uncertainty is estimated using 50 mock spectra created by randomly adding Gaussian noise at each pixel with its scale equal to the spectral error at that pixel (e.g., \citealt{Shen19,Wang20a,Yang20}). All the $1~\sigma$ uncertainties are then estimated based on the 16\% and 84\% percentile of the distribution. The bolometric luminosity [$L_{\rm bol}=(5.17\pm0.21)\times10^{47}\rm~erg~s^{-1}$] is estimated from the continuum luminosity at $3000\rm~\AA$ [$L_{\rm 3000 \AA}=(3.34\pm0.14)\times10^{43}\rm~erg~s^{-1}~\AA^{-1}$] with a bolometric correction factor of $\rm BC_{\rm 3000}=5.15$ \citep{Shen11}. The mass of the SMBH [($M_{\rm SMBH}=(1.82\pm0.02)\times10^9\rm~M_\odot$)] is then estimated based on $L_{\rm bol}$ and the FWHM of the \ion{Mg}{2} line [$(2817\pm22)\rm~km~s^{-1}$] by adopting the local empirical relation from \citet{Vestergaard09}. The uncertainty on $M_{\rm SMBH}$ estimated here does not include the systematic uncertainties of the scaling relation, which could be up to $\sim0.55\rm~dex$. We also estimate the corresponding Eddington ratio based on the measured $L_{\rm bol}$ and $M_{\rm SMBH}$, which is $\lambda_{\rm Edd}=2.25\pm0.09$.


\subsection{The Possible Short Timescale Variation of J074749+115352} \label{subsec:PeriodicJ074749}

The statistical significance of the X-ray variation of J074749+115352 depends on the time binning of the light curve. A lower time resolution could reduce the statistical uncertainty while may also reduce the amplitude of the variation, so could make the X-ray variation either more or less significant. At some time resolutions such as $\Delta t=2500\rm~s$, we could get a positive excess variance and a fractional rms variability amplitude of $F_{\rm var}\approx17\%$ \citep{Vaughan03}. However, considering the large uncertainties, we conclude that the apparent short timescale variation of J074749+115352 is not statistically significant based on the existing \emph{Chandra} observations.

Adopting the putative 3838~s period determined in \S\ref{Sec:ObsDataReduc}, the distinguish of the X-ray properties of the ``High'' and ``Low'' states is quite clear. As shown in Fig.~\ref{fig:J074749lc}d and listed in Table~\ref{table:XrayProperties}, even if there may be some mixture of photons from different states due to the poorly constrained variation period (also indicated in Fig.~\ref{fig:J074749lc}a), the average X-ray flux in the ``High'' state is still significantly higher than ($\sim2.7$ times of) that in the ``Low'' state ($F_{\rm 0.5-2 keV}=5.86_{-1.41}^{+1.36}\times10^{-14}\rm~erg~s^{-1}~cm^{-2}$ vs $2.20_{-0.43}^{+0.42}\rm~erg~s^{-1}~cm^{-2}$). Furthermore, the X-ray spectrum in the ``High'' state is also significantly softer ($\Gamma=2.40_{-0.32}^{+0.33}$ compared to $\Gamma=1.78_{-0.24}^{+0.25}$ in the low state). The distinguish of such ``high soft'' and ``low hard'' states is quite common in the X-ray variation of both the SMBHs and stellar mass BHs, which is thought to be mainly caused by the change of the accretion rate (e.g., \citealt{Nowak95,Done05}). But more data are need to monitor the long-term variation of J074749+115352 in order to better understand the transition of its spectral states. 

The short timescale variation detected in J074749+115352 is quite similar to the quasi-periodic oscillation (QPO) observed at low redshifts. Such a short timescale QPO is rarely detected in a quasar hosting such a massive SMBH, which typically shows much longer timescale X-ray variations (e.g., \citealt{McHardy06}). Using the scaling relations from \citet{GonzalezMartin12}, the expected variation period of a $\gtrsim10^9\rm~M_\odot$ SMBH (\S\ref{subsec:MSMBHJ074749}) should be about three orders of magnitude of what we have found in J074749+115352 ($\sim70\rm~days$ adopting its SMBH mass). Therefore, this apparent short timescale variation, if really exists, cannot be attributed to any ordinary orbital motions of the accretion disk. \citet{Gierlinski08} have discovered a $\sim1\rm~hour$ scale QPO in a $z=0.042$ narrow-line Seyfert~1 galaxy RE~J1034+396 (further confirmed by \citealt{Jin20a}), which hosts a SMBH with a poorly constrained mass of $\sim10^{5.8-7.6}\rm~M_\odot$. \citet{Jin20b} further decompose the X-ray emission into four components, including a disc component plus three components (two warm and one hot) from the corona. It is the hotter and less luminous warm corona component producing the QPO. The authors speculate that the QPO is due to the expansion/contraction of the inner disc vertical structure. A more sensitive X-ray telescope, such as the \emph{XMM-Newton}, is certainly needed for a more reliable spectral-timing analysis to confirm the nature of the X-ray variability of J074749+115352.

\subsection{Comparison to Other Quasar Samples} \label{subsec:ComparisonOthers}

Existing studies have shown a tight correlation between the rest frame UV luminosity of a quasar (typically $L_{\rm 2500 \AA}$) and its optical/UV-to-X-ray spectral slope $\alpha_{\rm OX}$ (e.g., \citealt{Just07,Lusso16}), which traces the relative importance of the emissions from the accretion disc and the corona. We herein compare the three quasars studied in this work to some other surveys of quasars at different redshifts.


\begin{figure}
\begin{center}
\epsfig{figure=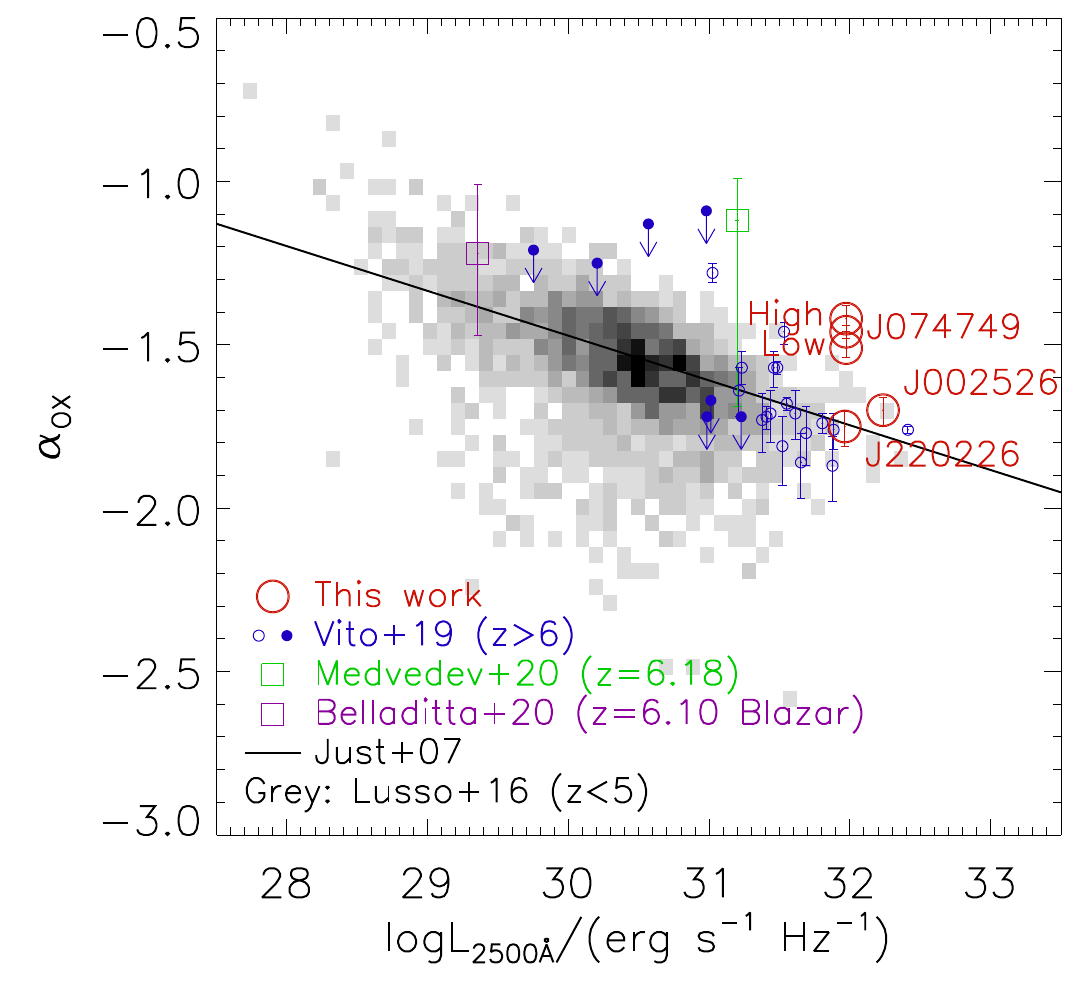,width=0.47\textwidth,angle=0, clip=}
\caption{Comparison of our $z\gtrsim5$ quasars with other quasar samples on the well defined $L_{\rm 2500 \AA}-\alpha_{\rm OX}$ relation. The three quasars studied in this work, as well as the ``High'' and ``Low'' states of J074749-115352 are plotted with large red open circles. The error bars are smaller than the size of the symbols. Blue dots are the $z>6$ quasars from \citet{Vito19}, with the open ones denote firm X-ray detections, while the solid ones denote upper limits on X-ray detections. The purple and green boxes are the recent discovery of two X-ray bright quasars at $z>6$ from \citet{Belladitta20} and \citet{Medvedev20}, respectively. The quasar from \citet{Medvedev20} is the X-ray brightest one known at $z>5$, while J074749+115352 from this work is the second X-ray brightest. Grey shaded pixels are a large sample of 2153 firmly X-ray detected quasars at $z<5$ from \citet{Lusso16}. The darkness of the pixel is proportional to the number of quasars in the corresponding $L_{\rm 2500 \AA}$ and $\alpha_{\rm OX}$ bins. We discarded their upper limits on X-ray non-detected sources to avoid confusion. The solid line is the best fit relation from \citet{Just07} (Fig.~7 in that paper). }\label{fig:L2500Aalphaox}
\end{center}
\end{figure}

There are two major samples plotted in Fig.~\ref{fig:L2500Aalphaox} together with the best-fit relation from \citet{Just07}. The \citet{Lusso16}'s sample is constructed by cross-matching the SDSS quasar sample with the 3XMM-DR5 catalog of X-ray sources. We only plot their firmly X-ray detected subsample in Fig.~\ref{fig:L2500Aalphaox}, which consists of 2153 quasars spreading in a broad redshift range of $z<5$. \citet{Lusso16}'s sample does not include higher-$z$ quasars which have been increasingly detected in X-ray in recent years (e.g., \citealt{Nanni17,Vito18,Pons19,Salvestrini19,Wang20b}). We therefore plot another high-$z$ quasar sample from \citet{Vito19} for comparison, which consists of only $z>6$ quasars and certainly biased to the most luminous ones. In addition to these two samples, we also plot two X-ray bright quasars at $z>6$ recently detected in X-ray, which have relatively flat optical/UV-to-X-ray slope (larger $\alpha_{\rm OX}$), similar as J074749+115352. PSO~J030947.49+271757.31 ($z=6.10$) is the first blazar and also the radio brightest AGN at $z>6$, with the X-ray emission detected with \emph{Swift} by \citet{Belladitta20}. CFHQS~J142952+544717 is the most distant known radio-loud quasar at $z=6.18$. Its X-ray emission was detected with \emph{eROSITA} by \citet{Medvedev20}. It is the X-ray brightest quasar at $z>5$ with an X-ray flux about twice of J074749+115352.

As shown in Fig.~\ref{fig:L2500Aalphaox}, two of the three quasars studied in this work, J002526-014532 and J220226+150952, are well consistent with the $L_{\rm 2500 \AA}-\alpha_{\rm OX}$ relation defined with low-$z$ quasars and most of the $z>6$ quasars (many of \citealt{Vito19}'s quasars with large $\alpha_{\rm OX}$ are upper limits). Their relatively high X-ray luminosity is only a result of their high UV luminosity. However, J074749+115352 appears to be extraordinarily X-ray bright at its UV luminosity, as indicated by the significantly larger $\alpha_{\rm OX}$ than the best-fit relation. Adopting a bolometric luminosity of $L_{\rm bol}\sim5\times10^{47}\rm~erg~s^{-1}$ roughly estimated from the $M_{\rm 1450}$, J074749+115352 is $\sim 4$ times as bright as expected from the $L_{\rm bol}-L_{\rm X}$ scaling relation (\citealt{Wang20b}; similar as predicted from a $L_{\rm UV}-L_{\rm X}$ relationship, e.g., \citealt{Salvestrini19}). 

X-ray observations from AGNs are expected to be suppressed compared to the UV luminosity when the Eddington ratio is high (e.g., \citealt{Jiang19}). Observations indicate an increasing bolometric correction factor in X-ray band (i.e., a higher bolometric to X-ray luminosity ratio) with the increasing Eddington ratio or bolometric luminosity (e.g., \citealt{Lusso12,Duras20}). We do not have a measurement of the SMBH mass and Eddington ratio for J002526-014532 and J220226+150952. Adopting the average rest frame 2-10~keV luminosity of $L_{\rm 2-10 keV}=(1.22\pm0.16)\times10^{46}\rm~erg~s^{-1}$ (Table~\ref{table:XrayProperties}) and the bolometric luminosity of $L_{\rm bol}=(5.17\pm0.21)\times10^{47}\rm~erg~s^{-1}=(1.35\pm0.05)\times10^{14}\rm~L_\odot$ for J074749+115352 (\S\ref{subsec:MSMBHJ074749}), we obtain a 2-10~keV bolometric correction factor of $L_{\rm bol}/L_{\rm2-10keV}=42.4\pm5.8$. At the Eddington ratio of $\lambda_{\rm Edd}\sim2$ or the bolometric luminosity of $L_{\rm bol}\sim10^{14}\rm~L_\odot$ for J074749+115352, the expected X-ray bolometric correction factor is typically $L_{\rm bol}/L_{\rm2-10keV}>60$ based on the $L_{\rm bol}/L_{\rm2-10keV}-\lambda_{\rm Edd}$ relationship from \citet{Lusso12}, or $L_{\rm bol}/L_{\rm2-10keV}>100$ based on the $L_{\rm bol}/L_{\rm2-10keV}-L_\odot$ relationship from \citet{Duras20}. J074749+115352 is radio quiet so the strong X-ray emission is unlikely produced by the jet \citep{Yang16}. We therefore conclude that this quasar is extraordinarily X-ray bright compared to most of the AGNs.

As an important probe of the accretion physics, there are a few studies of the redshift evolution of the X-ray spectral slope of high-$z$ quasars (e.g., \citealt{Vito19,Wang20b}). Although there are still very few measurement of $\Gamma$ for high-$z$ quasars ($\lesssim20$ for $z>5$ quasars), we notice that a constant X-ray spectral slope of $\Gamma\approx1.9$ at $z<6$ has been claimed, with a steeper average X-ray spectral slope (larger $\Gamma$) at $z\gtrsim6$. The increase of $\Gamma$ at $z\gtrsim6$, however, is probably caused by the higher $\lambda_{\rm Edd}$ of the $z\gtrsim6$ quasar samples \citep{Wang20b}. We add reliable X-ray measurements of $\Gamma$ of three quasars into the sample of $z\sim5$ quasars. The X-ray spectral slope of our quasars (the best-fit $\Gamma$ is in the range of 1.6-2.1) is consistent with the average $\Gamma$ of $z=4-6$ quasars, considering the error of the measurement and the dependence on the unknown $\lambda_{\rm Edd}$ (currently only J074749+115352 has an estimate of $\lambda_{\rm Edd}$). We therefore do not find a significant redshift evolution of $\Gamma$ based on the existing X-ray data.

\section{Summary} \label{Sec:Summary}

In this paper, we report new \emph{Chandra} observations of three $z\sim5$ quasars, which are the UV brightest radio quiet ones at the corresponding redshift. Significant X-ray emissions have been clearly detected and a high quality X-ray spectrum can be extracted to measure the basic X-ray properties of each quasar. 
All of the three quasars have a rest frame 2-10~keV luminosity $L_{\rm 2-10 keV}>10^{45}\rm~erg~s^{-1}$ and the X-ray power law photon index well constrained ($\Gamma$ in the range of 1.6-2.1). In particular, we confirm that J074749+115352 is the X-ray brightest radio quiet quasar at $z>5$ with an average observed X-ray flux $F_{\rm 0.5-2 keV}=3.49_{-0.45}^{+0.44}\times10^{-14}\rm~erg~s^{-1}~cm^{-2}$ (or a mean \emph{Chandra}/ACIS-S counts rate $cps_{\rm 0.5-7~keV}\approx4.3\times10^{-3}\rm~counts~s^{-1}$).

The high X-ray flux of J074749+115352 makes it unique for timing analysis on a timescale of a few hours at such high redshift. We find that J074749+115352 may have some short timescale variations, although such variations may be embedded in the large statistical noise. The best-fit period in the observer's frame is $P_{\rm obs}\approx 3838\rm~s$ with a corresponding rest frame period of $P_{\rm rest}\approx 613\rm~s$. We extract period folded light curve and phase folded spectra, and find that there are two clearly distinguishable states in the X-ray variation: a ``high soft'' state with an average X-ray flux $\sim2.7$ times of the ``low hard'' state, and the X-ray spectral slope is also significantly steeper ($\Gamma=2.40_{-0.32}^{+0.33}$ v.s. $1.78_{-0.24}^{+0.25}$). We also estimate the mass of the SMBH in J074749+115352 based on a near-IR spectrum covering the \ion{Mg}{2}~$\lambda2800\rm~\AA$ emission line, which is $M_{\rm SMBH}=(1.82\pm0.02)\times10^9\rm~M_\odot$. This is the first time to detect possible X-ray variation on a timescale of a few ks around such a massive SMBH at such a high redshift, although more sensitive X-ray observations are needed to examine this short timescale X-ray variation.

We also compare the three quasars detected in this paper to other quasar samples. We find that J074749+115352 is extraordinarily X-ray bright, with an average $\alpha_{\rm OX}=-1.46\pm0.02$ and 2-10~keV bolometric correction factor $L_{\rm bol}/L_{\rm2-10keV}=42.4\pm5.8$, both significantly depart from some well defined scaling relations. This quasar also has a high Eddington ratio of $\lambda_{\rm Edd}=2.25\pm0.09$. More X-ray and IR observations are needed to confirm the nature and better understand the properties of this unique quasar at the end of the cosmic dawn. This work add reliable X-ray measurements of $\Gamma$ of three quasars at $z\sim5$, but we do not find a significant redshift evolution of $\Gamma$ based on the existing limited sample of high-$z$ quasars with high-quality X-ray data.

\acknowledgments
The authors would like to acknowledge the anonymous referee for helpful comments and suggestions. We acknowledge Dr. Abderahmen Zoghbi at the University of Michigan for helpful discussions on the timing analysis of J074749+115352. The authors also acknowledge the use of data obtained at the Palomar 5.1m telescope. JTL acknowledge the financial support of this project from NASA and the Smithsonian Institution through the grant GO0-21097X, as well as the general financial support from NASA through the grant 80NSSC19K0579 and from the Smithsonian Institution through the grants GO9-20074X and AR9-20006X.
FW thanks the support provided by NASA through the NASA Hubble Fellowship grant \#HST-HF2-51448.001-A awarded by the Space Telescope Science Institute, which is operated by the Association of Universities for Research in Astronomy, Incorporated, under NASA contract NAS5-26555.

\end{document}